
\bigskip\bigskip\bigskip\bigskip\banner
\vskip 3 truecm
\footnoterule
\noindent$^1$ \AZ\par
\noindent$^2$ \BI\par
\noindent$^3$ \PENN\par
\noindent$^*$ Address after 1 October: DAMTP, University of Liverpool,
Liverpool L69 3BX, UK.\par
\noindent$^{**}$ Address after 1 October: CERN, Theory Division, CH-1211
Geneva 23, Switzerland.
\vfill\eject
\input phasedef
\input phase1
\input phase2
\input phase3
\input phase4
\input phase5
\input phaseref
\input phasef
\vfill\eject
\bye

\magnification=1200\overfullrule=0pt\baselineskip=15pt
\vsize=22truecm \hsize=15truecm \overfullrule=0pt\pageno=0

\font\titlefont=cmbx10 scaled \magstep1
\font\sectnfont=cmbx8  scaled \magstep1
\def\mname{\ifcase\month\or January \or February \or March \or April
           \or May \or June \or July \or August \or September
           \or October \or November \or December \fi}
\def\date{\hbox{\strut\mname \number\year}}

\def\binum{\hbox{BI-TP 93/37\strut}}
\def\banner{\hfill\hbox{\vbox{\offinterlineskip
                                \binum\date}}\relax}

\footline={\ifnum\pageno=0{}\else\hfil\number\pageno\hfil\fi}
%
%
%
%
\newcount\REFERENCENUMBER\REFERENCENUMBER=0
\def\REF#1{\expandafter\ifx\csname RF#1\endcsname\relax
               \global\advance\REFERENCENUMBER by 1
               \expandafter\xdef\csname RF#1\endcsname
                         {\the\REFERENCENUMBER}\fi}
\def\reftag#1{\expandafter\ifx\csname RF#1\endcsname\relax
               \global\advance\REFERENCENUMBER by 1
               \expandafter\xdef\csname RF#1\endcsname
                      {\the\REFERENCENUMBER}\fi
             \csname RF#1\endcsname\relax}
\def\ref#1{\expandafter\ifx\csname RF#1\endcsname\relax
               \global\advance\REFERENCENUMBER by 1
               \expandafter\xdef\csname RF#1\endcsname
                      {\the\REFERENCENUMBER}\fi
             [\csname RF#1\endcsname]\relax}
\def\refto#1#2{\expandafter\ifx\csname RF#1\endcsname\relax
               \global\advance\REFERENCENUMBER by 1
               \expandafter\xdef\csname RF#1\endcsname
                      {\the\REFERENCENUMBER}\fi
           \expandafter\ifx\csname RF#2\endcsname\relax
               \global\advance\REFERENCENUMBER by 1
               \expandafter\xdef\csname RF#2\endcsname
                      {\the\REFERENCENUMBER}\fi
             [\csname RF#1\endcsname--\csname RF#2\endcsname]\relax}
\def\refand#1#2{\expandafter\ifx\csname RF#1\endcsname\relax
               \global\advance\REFERENCENUMBER by 1
               \expandafter\xdef\csname RF#1\endcsname
                      {\the\REFERENCENUMBER}\fi
           \expandafter\ifx\csname RF#2\endcsname\relax
               \global\advance\REFERENCENUMBER by 1
               \expandafter\xdef\csname RF#2\endcsname
                      {\the\REFERENCENUMBER}\fi
            [\csname RF#1\endcsname,\csname RF#2\endcsname]\relax}
\def\refthree#1#2#3{\expandafter\ifx\csname RF#1\endcsname\relax
               \global\advance\REFERENCENUMBER by 1
               \expandafter\xdef\csname RF#1\endcsname
                      {\the\REFERENCENUMBER}\fi
           \expandafter\ifx\csname RF#2\endcsname\relax
               \global\advance\REFERENCENUMBER by 1
               \expandafter\xdef\csname RF#2\endcsname
                      {\the\REFERENCENUMBER}\fi
           \expandafter\ifx\csname RF#3\endcsname\relax
               \global\advance\REFERENCENUMBER by 1
               \expandafter\xdef\csname RF#3\endcsname
                      {\the\REFERENCENUMBER}\fi
           [\csname RF#1\endcsname, \csname RF#2\endcsname,
            \csname RF#3\endcsname]\relax}
\def\reftothree#1#2#3{\expandafter\ifx\csname RF#1\endcsname\relax
               \global\advance\REFERENCENUMBER by 1
               \expandafter\xdef\csname RF#1\endcsname
                      {\the\REFERENCENUMBER}\fi
           \expandafter\ifx\csname RF#2\endcsname\relax
               \global\advance\REFERENCENUMBER by 1
               \expandafter\xdef\csname RF#2\endcsname
                      {\the\REFERENCENUMBER}\fi
           \expandafter\ifx\csname RF#3\endcsname\relax
               \global\advance\REFERENCENUMBER by 1
               \expandafter\xdef\csname RF#3\endcsname
                      {\the\REFERENCENUMBER}\fi
            [\csname RF#1\endcsname--\csname RF#3\endcsname]\relax}
\def\reffour#1#2#3#4{\expandafter\ifx\csname RF#1\endcsname\relax
               \global\advance\REFERENCENUMBER by 1
               \expandafter\xdef\csname RF#1\endcsname
                      {\the\REFERENCENUMBER}\fi
           \expandafter\ifx\csname RF#2\endcsname\relax
               \global\advance\REFERENCENUMBER by 1
               \expandafter\xdef\csname RF#2\endcsname
                      {\the\REFERENCENUMBER}\fi
           \expandafter\ifx\csname RF#3\endcsname\relax
               \global\advance\REFERENCENUMBER by 1
               \expandafter\xdef\csname RF#3\endcsname
                      {\the\REFERENCENUMBER}\fi
           \expandafter\ifx\csname RF#4\endcsname\relax
               \global\advance\REFERENCENUMBER by 1
               \expandafter\xdef\csname RF#4\endcsname
                      {\the\REFERENCENUMBER}\fi
            [\csname RF#1\endcsname, \csname RF#2\endcsname,
             \csname RF#3\endcsname, \csname RF#4\endcsname]\relax}
\def\reftofour#1#2#3#4{\expandafter\ifx\csname RF#1\endcsname\relax
               \global\advance\REFERENCENUMBER by 1
               \expandafter\xdef\csname RF#1\endcsname
                      {\the\REFERENCENUMBER}\fi
           \expandafter\ifx\csname RF#2\endcsname\relax
               \global\advance\REFERENCENUMBER by 1
               \expandafter\xdef\csname RF#2\endcsname
                      {\the\REFERENCENUMBER}\fi
           \expandafter\ifx\csname RF#3\endcsname\relax
               \global\advance\REFERENCENUMBER by 1
               \expandafter\xdef\csname RF#3\endcsname
                      {\the\REFERENCENUMBER}\fi
           \expandafter\ifx\csname RF#4\endcsname\relax
               \global\advance\REFERENCENUMBER by 1
               \expandafter\xdef\csname RF#4\endcsname
                      {\the\REFERENCENUMBER}\fi
            [\csname RF#1\endcsname--\csname RF#4\endcsname]\relax}
\def\reftosix#1#2#3#4#5#6{\expandafter\ifx\csname RF#1\endcsname\relax
               \global\advance\REFERENCENUMBER by 1
               \expandafter\xdef\csname RF#1\endcsname
                      {\the\REFERENCENUMBER}\fi
           \expandafter\ifx\csname RF#2\endcsname\relax
               \global\advance\REFERENCENUMBER by 1
               \expandafter\xdef\csname RF#2\endcsname
                      {\the\REFERENCENUMBER}\fi
           \expandafter\ifx\csname RF#3\endcsname\relax
               \global\advance\REFERENCENUMBER by 1
               \expandafter\xdef\csname RF#3\endcsname
                      {\the\REFERENCENUMBER}\fi
           \expandafter\ifx\csname RF#4\endcsname\relax
               \global\advance\REFERENCENUMBER by 1
               \expandafter\xdef\csname RF#4\endcsname
                      {\the\REFERENCENUMBER}\fi
           \expandafter\ifx\csname RF#5\endcsname\relax
               \global\advance\REFERENCENUMBER by 1
               \expandafter\xdef\csname RF#5\endcsname
                      {\the\REFERENCENUMBER}\fi
           \expandafter\ifx\csname RF#6\endcsname\relax
               \global\advance\REFERENCENUMBER by 1
               \expandafter\xdef\csname RF#6\endcsname
                      {\the\REFERENCENUMBER}\fi
            [\csname RF#1\endcsname--\csname RF#6\endcsname]\relax}
\def\reftofive#1#2#3#4#5{\expandafter\ifx\csname RF#1\endcsname\relax
               \global\advance\REFERENCENUMBER by 1
               \expandafter\xdef\csname RF#1\endcsname
                      {\the\REFERENCENUMBER}\fi
           \expandafter\ifx\csname RF#2\endcsname\relax
               \global\advance\REFERENCENUMBER by 1
               \expandafter\xdef\csname RF#2\endcsname
                      {\the\REFERENCENUMBER}\fi
           \expandafter\ifx\csname RF#3\endcsname\relax
               \global\advance\REFERENCENUMBER by 1
               \expandafter\xdef\csname RF#3\endcsname
                      {\the\REFERENCENUMBER}\fi
           \expandafter\ifx\csname RF#4\endcsname\relax
               \global\advance\REFERENCENUMBER by 1
               \expandafter\xdef\csname RF#4\endcsname
                      {\the\REFERENCENUMBER}\fi
           \expandafter\ifx\csname RF#5\endcsname\relax
               \global\advance\REFERENCENUMBER by 1
               \expandafter\xdef\csname RF#5\endcsname
                      {\the\REFERENCENUMBER}\fi
            [\csname RF#1\endcsname--\csname RF#5\endcsname]\relax}
\def\Ref#1{\expandafter\ifx\csname RF#1\endcsname\relax
               \global\advance\REFERENCENUMBER by 1
               \expandafter\xdef\csname RF#1\endcsname
                      {\the\REFERENCENUMBER}\fi
             Ref.~[\csname RF#1\endcsname]\relax}
\newcount\EQUATIONNUMBER\EQUATIONNUMBER=0
\def\EQ#1{\expandafter\ifx\csname EQ#1\endcsname\relax
               \global\advance\EQUATIONNUMBER by 1
               \expandafter\xdef\csname EQ#1\endcsname
                          {\the\EQUATIONNUMBER}\fi}
\def\eqtag#1{\expandafter\ifx\csname EQ#1\endcsname\relax
               \global\advance\EQUATIONNUMBER by 1
               \expandafter\xdef\csname EQ#1\endcsname
                      {\the\EQUATIONNUMBER}\fi
            \csname EQ#1\endcsname\relax}
\def\EQNO#1{\expandafter\ifx\csname EQ#1\endcsname\relax
               \global\advance\EQUATIONNUMBER by 1
               \expandafter\xdef\csname EQ#1\endcsname
                      {\the\EQUATIONNUMBER}\fi
            \eqno(\csname EQ#1\endcsname)\relax}
\def\EQNM#1{\expandafter\ifx\csname EQ#1\endcsname\relax
               \global\advance\EQUATIONNUMBER by 1
               \expandafter\xdef\csname EQ#1\endcsname
                      {\the\EQUATIONNUMBER}\fi
            (\csname EQ#1\endcsname)\relax}
\def\eq#1{\expandafter\ifx\csname EQ#1\endcsname\relax
               \global\advance\EQUATIONNUMBER by 1
               \expandafter\xdef\csname EQ#1\endcsname
                      {\the\EQUATIONNUMBER}\fi
          Eq.~(\csname EQ#1\endcsname)\relax}
%
\newcount\SECTIONNUMBER\SECTIONNUMBER=0
\newcount\SUBSECTIONNUMBER\SUBSECTIONNUMBER=0
\def\section#1{\global\advance\SECTIONNUMBER by 1\SUBSECTIONNUMBER=0
      \bigskip\goodbreak\line{{\bf \the\SECTIONNUMBER.\ #1}\hfil}
      \smallskip}
\def\subsection#1{\global\advance\SUBSECTIONNUMBER by 1
      \bigskip\goodbreak\line{{\sectnfont
         \the\SECTIONNUMBER.\the\SUBSECTIONNUMBER.\ #1}\hfil}
      \smallskip}
%
%

%
%
\def\E{{\scriptscriptstyle E}}
%

%
\def\avg#1{\ifmmode\langle#1\rangle\else$\langle#1\rangle$\fi}
\def\X{\overline x}\def\x{\ifmmode\X\else$\X$\fi}
\def\MUE{\mu_\E}\def\mue{\ifmmode\MUE\else$\MUE$\fi}

%
%
%
%
\begingroup\titlefont\obeylines
\hfil The Physical Phase of
\hfil Dimensionally Reduced Gauge Theories \hfil
\endgroup\bigskip
\def\footnoterule{\kern-3pt
                  \hrule width 2 true in
                  \kern 2.6pt}
\def\BI{Fakult\"at f\"ur Physik, Universit\"at Bielefeld,
Postfach 100131, 33501 Bielefeld, Germany\hfill}
\def\PENN{Physics Department, Penn State University,
Hazleton, Pennsylvania 18201, USA.\hfill}
\def\AZ{Physics Department, University of Arizona,
Tucson AZ 85721, USA.\hfill}
\medskip
\centerline{L.~K\"arkk\"ainen$^1$,
P. Lacock$^{2*}$,
D.E. Miller$^{2,3}$}
\centerline{B. Petersson$^2$ and T. Reisz$^{2**}$}
\bigskip\bigskip\bigskip
\bigskip\bigskip\bigskip\centerline{{\sectnfont ABSTRACT}}\medskip
We investigate the relationship between the high temperature
deconfined phase of the SU(2) gauge theory to the phases of the
corresponding three dimensional adjoint Higgs model.
For various temperatures we simulate the effective theory
in a neighbourhood of the physical states, that is of
those values of the coupling constants that describe the
infrared behaviour of the four dimensional theory and which
have been calculated by applying dimensional reduction
techniques.
We show that the physical points belong to the confined phase
of the SU(2) adjoint Higgs model.

\bigskip\bigskip\bigskip\bigskip\banner
\vskip 3 truecm
\footnoterule
\noindent$^1$ \AZ\par
\noindent$^2$ \BI\par
\noindent$^3$ \PENN\par
\noindent$^*$ Address after 1 October: DAMTP, University of Liverpool,
Liverpool L69 3BX, UK.\par
\noindent$^{**}$ Address after 1 October: CERN, Theory Division, CH-1211
Geneva 23, Switzerland.
\vfill\eject
\vfill\eject

\def\O{{\cal O}}

\def\O{{O}}

\def\tr{{\rm tr}}

\def\lsim{\raise 0.3ex\hbox{
     $<$\kern -0.75em\raise -1.1ex\hbox{$\sim$}}}
\def\gsim{\raise 0.3ex\hbox{
     $>$\kern -0.75em\raise -1.1ex\hbox{$\sim$}}}

%
\section{Introduction}
\medskip
In recent works \reftofour{thomas}{su2}{su3}{qcd}
we have shown how SU($N_c$) gauge theories and
QCD in
four dimensions (4D) at finite temperature reduce at high
temperature to SU$(N_c)$ adjoint Higgs models in three
dimensions (3D). In particular we made a detailed analysis
of the process of dimensional reduction in the electrostatic
sector
in a temperature range between two and eight
times the critical temperatures of these theories - in all cases
clearly in the deconfined phase of the 4D gauge theory. Nevertheless,
we were at the time unable to provide
any clearcut evidence concerning the phase of the effective
system based on previous knowledge of the phase structure for the
related 3D adjoint Higgs models. In this work we explicitly address
the relationship of the deconfined phase of the SU(2) gauge theory
in 4D to the phases of an SU(2) adjoint Higgs model in 3D.

It has been known \reftothree{mcl}{kuti}{engels}
for a long time that in 4D the pure
SU(2) gauge theory at finite temperature undergoes a second order
phase transition at a critical temperature around 200 MeV.
The simple physical picture has the gluons essentially free for
temperatures above this critical temperature so that they are said
to be in a deconfined phase.

It has been claimed that the infrared behaviour of the
deconfined phase is described by the Higgs phase of the effective
SU(2) Higgs model.
Although the phases of this model
have been well studied in 4D \refand{baier1}{baier2}, this
situation is not carried over to the same extent in 3D where it
serves as the effective theory \reftothree{n1}{n2}{n3}.
It has been pointed out \ref{n4} that in 3D
the effective model has a very special phase structure, which is
dominated by a "confined" (SU(2) symmetric)
and a "Higgs" (U(1) symmetric) phase,
which are analytically connected.
Long ago Polyakov \ref{poly} argued by semiclassical approximations
that the infrared behaviour in the
Higgs phase can be understood as a dilute Coulomb gas of
't Hooft-Polyakov
monopoles with long range interactions which may be
seen as a disordered system. However the broken $SU(2)$
symmetry is partially restored with a U(1) symmetric phase.
This monopole system produces a U(1)
confinement, which explains the apparently contradictory requirements
of having both confining and Higgs behaviour.

In this work we shall present our results on the actual phase
realised in the effective 3D model that is derived by
dimensional reduction techniques \refand{thomas}{su2}.
The following section
includes some known properties of the effective model as well as a
discussion of its phase structure.
In the third section we discuss the numerical methods employed
here while in the fourth section we present and discuss
our numerical results. Finally we state our conclusions.

\section{The Effective Model and the Phases}
\medskip

In our previous analyses \reftofour{thomas}{su2}{su3}{qcd}
of the process of dimensional reduction from 4D to 3D
we have made explicit the form of the
effective action and the values of its coupling constants
relevant for the high temperature behaviour
at least of the electric sector of the plasma phase.
On the lattice $\Lambda$ the basic fields are the SU(2) valued link
variables $U(\vec x; i)$ and the adjoint scalar field
$$
   A_0 (\vec x) = i \sum^3_{a=1} A^a_0 ( \vec x ) \sigma^a,~~~\vec x \in
   \Lambda, \quad A_0^a \;{\rm real} , \EQNO{2.1}
   $$
where $\sigma^a$ are the Pauli matrices.
Thus we write the
effective action in the following form \ref{su2}:
$$S_{eff} (U, A_0) = S_w (U) + S_{hop} (U, A_0) + S_{sc} (A_0)
\EQNO{2.2}$$
where $S_w (U)$ is the standard 3D Wilson action, $S_{hop} (U, A_0)$
denotes the kinetic hopping term
$$S_{hop} (U, A_0) = {1\over 2} \beta \sum_{\vec x} \sum^3_{i=1}
\tr (A_0 (\vec x) U (\vec x ; i) A_0 ( \vec x + i) U(\vec x ; i)^{-1}),
\EQNO{2.3}$$
while $S_{sc} (A_0)$ is the purely scalar part of the action
$$S_{sc} (A_0) = {1\over 2} \beta \sum_{\vec x} ( - (3 + {1\over 2} h)
\;\tr A^2_0 (\vec x) +  K     ({1\over 2} \;\tr A^2_0 (\vec x ))^2).
\EQNO{2.4}$$
The parameters of our effective model are $\beta$, which relates
to the 3D gauge coupling,
and the quadratic and quartic scalar self couplings
$h$ and $ K    $, respectively.
They have been calculated explicitly by means of renormalised
perturbation theory
and are listed in Table 1 for various temperatures.
These three parameters differ from the usual three parameters
\reftosix{baier1}{baier2}{n1}{n2}{n3}{n4}
of the SU(2) adjoint Higgs model, which are usually written as
$\beta,~\kappa$ and $\lambda$, whereby $\beta$ remains the same and
$\kappa$ and $\lambda$ may be related to our parameters by simple
algebraic equations.
In terms of our parameters \reftothree{su2}{su3}{qcd}
we represent the usual ones
as follows:
$$\eqalign{
  {1 - 2 \lambda \over \kappa } \; & = \;
    3 + {h \over 2} ,  \cr
  {\lambda \over \kappa^2 } \; & = \;
    {K \over 2 \beta } \; . \cr}  \EQNO{2.5}
    $$
It turns out that our parametrisation is the more convenient
one for the localisation of the physical states
relative to the phase transition line.
In particular, $h$ is the mass term parameter of the scalar field.

Let us summarise what is known about the phase structure
of adjoint Higgs models.
The phase structure of the SU(2) adjoint
Higgs model has been generally well studied. In 4D the phase diagram
of this model in the parameter space $(\beta, \kappa, \lambda)$
has been constructed by applying the mean field and Monte Carlo
renormalisation group methods
\refand{baier1}{baier2}. This
work in 4D shows a multitude of first and second order phase
transitions with various branchings of the boundary lines depending
upon the parameters. It is clear that at small values of all these
parameters a strong coupling SU(2) confined phase is present. In the
case of large values of these parameters the Higgs field with a
fixed length $(\lambda = \infty)$ has been well studied. Here a
transition is found which connects the U(1) gauge theory in the
large $\kappa$ limit to the O(3) spin model in the weak coupling
limit \ref{baier2}. Thus in 4D the Higgs phase at large values of the
couplings is well separated from the confined phase.

On the other hand it has been
stated by Nadkarni \ref{n4}
that in 3D this sort of simple separation of the
phases does not take place. The phases are analytically connected.
This situation for the purpose of dimensional reduction
calls for a more careful analysis of
the related phases in 3D and 4D.
The generally presented picture for 3D \ref{n4}
has a first order transition line between the confined and
Higgs phases at small values of $\lambda$
going over to a second order one when both $\lambda$ and $\beta$
become large. At small values of $\beta$ there is a
critical point at which the first order transition terminates.
This situation is quite different from the 4D adjoint Higgs
model \ref{baier1} where a first order transition further
separates the phases in the $\kappa$-$\lambda$ plane.

The main question we are concerned with here is to which
phase the physical states belong.
These states are described by the parameter values
of Table 1 for three different temperatures.
The temperature is encoded in the 4D coupling $\beta_4$
that corresponds to the $L_0\times L_s^3$ lattice
with $L_0=4$,
ranging from $\beta_4 = 2.50$ to
$\beta_4 = 3.00$.
The particular values of the effective coupling constants
are the representatives on the
"curve of constant physics" that correspond to the finite
lattice cutoff $a(\beta )^{-1} = 4T$, cf. e.g.
\ref{qcd}, and have been calculated in \ref{su2}.\par

It has been widely claimed that the effective Higgs phase should
correspond to the high temperature plasma phase
\refand{n4}{polonyi}.
A preliminary discussion, however, done in connection with the
derivation of the effective model, shows the following situation
\ref{su2}: using the location of
the first order phase transition between
the confined and the Higgs phase as given by Nadkarni
\ref{n4}
(based on the one loop effective potential without Higgs loop),
for fixed temperature the critical point
($\lambda = 0$, $\kappa = 1/3$, $\beta = \infty$) is
approached in the continuum limit from the SU(2) symmetric phase!
Already the coupling constants of Table 1 that correspond
to $a(\beta )=(4T)^{-1}$ appear to be close to the transition line,
but definitely on the side of the confined phase.\par

\vskip 1cm
{Table 1. The physical coupling constants of the effective
model for the various lattice sizes and temperatures considered here.}
$$\vbox{\halign{
\hfil#\quad\hfil &\hfil#\quad\hfil &
\hfil#\quad\hfil &\hfil#\quad\hfil &
#\hfil\quad&#\hfil\cr
\multispan6\hrulefill\cr
$\beta_4$ & $L_s$ & \hfil $\beta$ \hfil &
\hfil $h$ \hfil & \hfil $K$ \hfil &
\hfil $T/T_c$ \hfil \cr
\multispan6\hrulefill\cr
2.50 & 16 & 12.35 & -0.284 &
0.105 & 2.0 \cr
2.80 & 12 & 14.34 & -0.237 &
0.085 & 4.5 \cr
\qquad & 16 & 13.70 & -0.250 &
0.092 & \quad \cr
\qquad & 24 & 13.54 & -0.260 &
0.094 & \quad \cr
3.00 & 16 & 14.60 & -0.240 &
0.086 & 8.0 \cr
\multispan6\hrulefill\cr}}$$

Furthermore, the computation of the one loop Higgs potential
along the lines proposed in \ref{polonyi}
gives evidence that actually the symmetric phase with
vanishing Higgs condensate is realised in the effective
theory.
In order to confirm this statement, we need to investigate a
quantity that is sensitive to the phase transition.
In the unitary gauge
$$\eqalign{
    & A^1(\vec x) = A^2(\vec x) = 0 , \cr
    & A^3(\vec x) \geq 0 , \cr}
  \EQNO{2.6}$$
a natural order parameter might be the length of the Higgs
field.
Instead of gauge rotating each configuration we prefer to use
the gauge invariant length of the adjoint scalar field.
Because of
$$
   \det A_0(\vec x) = - {\rm tr}\; A_0^2(\vec x) \geq 0 ,
  $$
it is given by
$$
  \O = {1\over |\Lambda|} \sum_x \sqrt{\det A_0 (\vec x)} ,
  \EQNO{2.7}
  $$
with
$|\Lambda|$ the volume of the finite 3D lattice $\Lambda$.
Although it does not vanish in the confined phase,
it will provide a clear signal of the phase transition.

We also define the
two-point susceptibility
$$\chi = |\Lambda| \; ( < \O^2 > - < \O >^2 ) .
  \EQNO{chi} $$

\section{Numerical Methods}

\medskip

For the Monte Carlo update of the system described by the
action \eq{2.2}, we use the following methods:
the gauge fields are generated by means of a standard heat bath
algorithm and then accepted with the conditional weight
exp($-\Delta S_{hop}) > r$, where $r$ is some random number in
the interval between zero and one. For the scalar fields we
use a pure Metropolis update, keeping the gauge fields fixed.
The acceptance of the scalar fields is tuned to obtain
a value of around 35 \%, which implies that the acceptance
for the gauge fields is about 90 \% \ref{su2}.\par

In the next section we shall investigate three different temperature
($\beta_4$) couplings (see Table 1). The finite-size effects will
be investigated by considering, for fixed corresponding
4D coupling $\beta_4$, the behaviour of the phase
transition on two lattices of different sizes.
The lattice sizes we investigate in detail are $12^3$ and $16^3$.
We have also performed runs on a $24^3$ lattice for
$\beta_4 = 2.80$, but due to the large lattice volume were
unable to perform a reweighting analysis since the energy
histograms of the individual runs (for the different $h$-couplings)
no longer sufficiently overlap. To investigate volumes
of this size and larger would necessitate the use of more
efficient update schemes, e.g. the multicanonical algorithm
\ref{berg}. This was not attempted here, since we were
able to obtain clear signals of the phase transition
already on the two smaller lattices.

On the $16^3$ lattice we have in general performed simulations
consisting of around 80000 runs per
quadratic scalar self coupling $h$, while
on the $12^3$ lattice we have at least 300000 runs per coupling.
Measurements are taken as a rule every 10th sweep.
In addition to the correlations
of the scalar fields we also measure the values of
the different terms in the action as well as the value
of the order parameter $\O$. This enables us to make use
of well-known reweighting techniques \ref{fs} to obtain
continuous curves, e.g. for the order parameter as a function
of the coupling $h$. The method
employed here does not rely on binning the data sample,
but instead uses all the data in the reweighting \ref{kari}.\par

In the reweigthing analysis we take into account the
integrated correlation times of the measurements, which for
the scalar mass term (the term proportional to $h$ in \eq{2.4})
typically ranges from $\tau_{int} \approx 8$
away from the transition to more than a hundred at the transition
point itself. For couplings in the immediate vicinity of the
transition point we have to check that the system is properly
thermalised (due to the flip-flop behaviour).
The statistical errors are determined by means of a standard
jackknife procedure.

\section{Results}

\medskip

Our aim is to determine, for various temperatures $T$, the position
of the physical points in relation to the phase transition
in the 3D effective model.
The physical values that correspond to the lattice spacing
equal to $(4T)^{-1}$ are given by Table 1.
In order to locate the phase transition we keep
the parameters K and $\beta$ fixed and vary the value of $h$.
For small values of $|h|$ the system will be
in the confined phase, while for larger ones it should be in
the Higgs phase. In our evaluation we have taken three
temperatures ranging from two to eight times the critical temperature.
We consider the quantity $\O$, which is the length of the
adjoint Higgs field. This is not strictly speaking an order parameter
since its value does not range between zero and one. Nevertheless,
it shows a rapid rise at the transition to the Higgs phase.

In Figs. 1-3 we collect the results for the order parameter $\O$
as a function of the coupling $h$ for the different temperatures
under consideration.
Fig. 1 shows the Monte Carlo data points obtained from the
simulations on a $16^3$ lattice
that correspond to $\beta_4 = 2.80$ ($T/T_c \approx 4.5)$.
The solid curve is the result of the reweighting analysis
as stated in the previous section.
According to Table 1
the physical value of the coupling $h$, henceforth denoted by
$h_{ph}$, is given here by $h_{ph}=-0.250$ and
lies close to the phase transition
point, although still clearly within the confined phase.
We have explicitly checked this by performing runs with
both cold and hot starts ($ A_0(\vec x)$ = 0 or random respectively),
obtaining the same results in both cases.
Hence, for smaller values of $|h|$ including the
physical points we are in the
confined phase, while at larger values there is a
transition to the Higgs phase.
The results of the reweighting analysis for all three
temperatures are shown in Fig. 2.
In all cases, the behaviour of $\O$ suggests that
the phase transition has a first order nature, with
clear metastability and a very rapid transition
from one phase to the other.

To check the finite-size behaviour of the order parameter
as the lattice size is changed, we show in Fig. 3 the results
for $\beta_4= 2.80$ obtained on a $12^3$ lattice. The
quantity $\O$ behaves as expected for a decrease in the volume
-- the curve becomes more rounded compared to the one shown
in Fig. 1 for the $16^3$ lattice. However, we find that even
for the $12^3$ lattice the finite-volume effects seem to be
small, with $\O$ having a well-defined transitional behaviour
normally associated with larger volumes.

To further elucidate the position of the phase transition point and
the nature of the phase transition, we consider the
behaviour of the susceptibility (\eq{chi}).
In Fig. 4 we show the results for the susceptibility
normalised to the lattice volume
for $\beta_4 = 2.80$ obtained on
lattice sizes $12^3$ (dashed line) and $16^3$ (solid line).
We clearly see that in both cases the normalised susceptibility
has a well-located peak which, at the transition point, scales
approximately with the volume of the system ($\chi/V \approx$ constant).
The normalised susceptibility furthermore shows a
very narrow transition range for both lattice sizes, which
is even more pronounced for the larger lattice, as would
be expected from finite-size scaling \ref{borgs1}.

Although we have not performed the reweighting analysis for
the $24^3$ lattice, we can roughly deduce the transition
point by comparison of the behaviour of the MC data
(which again show hysteresis) with those
obtained on the two
smaller lattices. This value is listed in Table 2,
together with the values obtained on the other lattices.
The latter are inferred from the position of the peak in the
susceptibility in each instance.
It is clear that for all three lattice sizes
$|h_{tr}|$ grows with increasing lattice size
slightly faster than does the physical point $|h_{ph}|$.\par

\vskip 1cm
{Table 2. The quadratic scalar self-coupling. Compared
are the physical values that correspond to $\beta_4 = 2.80$ on
$4 \times L_s^3$ lattices ($T/T_c \approx 4.5$) according
to Table 1 and the transition points $h_{tr}$.}
$$\vbox{\halign{
\hfil#\quad\hfil &\hfil#\quad\hfil &
#\hfil\cr
\multispan3\hrulefill\cr
$ \hfil L_s$ \hfil &
\hfil $h_{ph}$ \hfil & \hfil $h_{tr}$ \hfil \cr
\multispan3\hrulefill\cr
12 & -0.237 & -0.2623(1) \cr
16 & -0.250 & -0.2803(1) \cr
24 & -0.260 & -0.30(1)~~ \cr
\multispan3\hrulefill\cr}}$$

In Fig. 5 we furthermore show the behaviour of the transition point
$h_{tr}$ as a function of the inverse lattice volume for
the runs corresponding to $\beta_4 = 2.80$.
The line is a best linear fit to the data. We see that the
critical couplings scale approximately with the inverse lattice
volume, which
is again in agreement with the expected behaviour for a
phase transition of first order.

We want to emphasise that there is no evidence
that the first order transition in the effective three dimensional
model in any way corresponds to the second order deconfinement
transition of the original four dimensional gauge theory. We always
stay in the confined phase of the effective model.
This is clearly shown in Fig. 6, where we see
no approach towards the Higgs phase transition
as the temperature decreases.



\section{Conclusions}

\medskip

The SU(2) adjoint Higgs model as described by the action
\eq{2.2} - \eq{2.4} has been extensively used to describe
the infrared behaviour of the electric sector of gauge theories
at high temperature, in particular for a quantitative
determination of the Debye screening mass.
The main result of the present work is that the effective
theory belongs to the SU(2) symmetric phase.
This is a result of both an investigation of the one loop
effective potential in the effective model
as computed in connection with \ref{su2},
and a detailed Monte Carlo calculation done in this work.
The gauge invariantly defined Higgs condensate turns out to
be the appropriate "order" parameter. Also, we stress
that the first order transition
in the effective three dimensional
model in no way corresponds to the second order deconfinement
transition of the original four dimensional gauge theory.

Finally, we want to make two remarks:
First, the fact that the effective model is in the SU(2) confined phase
should not be confused with magnetic confinement of the full theory.
For instance, a strong coupling expansion shows that spatial
Wilson loops always have an area law behaviour for sufficiently
high temperatures \ref{borgs2}.
However, in order to decide whether magnetic screening or confinement
holds in the 4d theory, one first of all needs to find the appropriate local
observables and their correlations.
For the effective theory this means that one then has to investigate
the corresponding effective operators.

Second, the effective coupling constants have been calculated
perturbatively.
It might happen that perturbation theory is
more slowly convergent for the magnetic observables than for the
electric ones. Thus it is desirable to push up the order one further
for the magnetic sector.
For instance, a magnetic mass term is missing in the effective
action to the one loop order, but could appear to the next order.
On the other hand, the effective model as considered here is
expected to generate a mass gap dynamically, so the influence
of such a (higher order) mass parameter
need not be of importance.
A detailed investigation by means of an infrared finite expansion
scheme (e.g. such as in \ref{pert})
should clarify the situation.

\bigskip
\noindent
{\bf Acknowledgements}
\medskip
One of us (D.E.M.) would like to thank Rudolf Baier and Krzysztof
Redlich for discussions as well as the Fakult\"at f\"ur Physik der
Universit\"at Bielefeld for a guest professorship.
We would like to acknowledge the computer time
at the Pittsburgh Supercomputer Center and HLRZ, J\"ulich.
This project is supported by the Deutsche Forschungsgemeinschaft and
U. S. Dept. of Energy grant No. DE-FG02-8SER40213.

\medskip

\bigskip{\bf References}\bigskip
\item{\reftag{thomas}.}T. Reisz, Z. f. Phys. $\underline{C53}$ (1992) 169.
\item{\reftag{su2}.} P. Lacock, D. E. Miller and T. Reisz, Nucl. Phys.
$\underline{B369}$ (1992) 501.
\item{\reftag{su3}.} L. K\"arkk\"ainen, P. Lacock, D. E. Miller,
B. Petersson
and T. Reisz, Phys. Lett. $\underline{B282}$ (1992) 121.
\item{\reftag{qcd}.} L. K\"arkk\"ainen, P. Lacock,
B. Petersson
and T. Reisz, Nucl. Phys. $\underline{B395}$ (1993) 733.
\item{\reftag{mcl}.} L. McLerran and B. Svetitsky, Phys. Lett.
$\underline{B98}$ (1981) 195.
\item{\reftag{kuti}.} J. Kuti, J. Polonyi and K. Szlachanyi, Phys. Lett.
$\underline{B98}$ (1981) 199.
\item{\reftag{engels}.} J. Engels, F. Karsch, I. Montvay and H. Satz,
Phys. Lett.
$\underline{B101}$ (1981) 89.
\item{\reftag{baier1}.} R. Baier and H.-J. Reusch,
Nucl. Phys. $\underline{B285}$
(1987) 535.
\item{\reftag{baier2}.} R. Baier, C.B. Lang and H.-J. Reusch,
Nucl. Phys. $\underline{B305}$
(1988) 396.
\item{\reftag{n1}.} S. Nadkarni, Phys. Rev. $\underline{D27}$ (1983) 917.
\item{\reftag{n2}.} S. Nadkarni, Phys. Rev. $\underline{D38}$ (1988) 3287.
\item{\reftag{n3}.} S. Nadkarni, Phys. Rev. Lett. $\underline{60}$ (1988) 491.
\item{\reftag{n4}.} S. Nadkarni, Nucl. Phys. $\underline{B334}$ (1990) 559.
\item{\reftag{poly}.}A. M. Polyakov, Nucl. Phys.$\underline{B120}$ (1977) 429.
\item{\reftag{polonyi}.}
J. Polonyi and S. Vazquez, Phys. Lett. $\underline{B240}$ (1990) 183.
\item{\reftag{berg}.} B. Berg and T. Neuhaus, Phys. Lett.
$\underline{B267}$ (1991) 249.
\item{\reftag{fs}.} A.M. Ferrenberg and R.H. Swendsen, Phys. Rev. Lett.
$\underline{61}$ (1988) 2635.
\item{\reftag{kari}.} K. Kajantie, L K\"arkk\"ainen and K. Rummukainen,
$\underline{B357}$ (1991) 693.
\item{\reftag{borgs1}.} C. Borgs and R. Kotecky, Jour. Stat. Phys.
$\underline{61}$ (1990) 79.
\item{\reftag{borgs2}.} C. Borgs, Nucl. Phys.
$\underline{B261}$ (1985) 455.
\item{\reftag{pert}.} B. Petersson, T. Reisz, Nucl. Phys.
$\underline{B353}$ (1991) 757.

\vfill\eject
\centerline{{\bf Figure captions}}\smallskip
\item{Fig. 1.} The order parameter $O$
as a function of the coupling -h on
$\beta_4 = 2.80$ for the $16^3$ lattice.
The circles and crosses represent,
respectively, $O$ for cold and hot starts obtained directly
from the MC runs. The solid curve is the result of the
reweighting analysis.
\item{Fig. 2.} A comparison between the reweighted
curves for $O$ on the $16^3$ lattice with the values
of $\beta_4$ equal to 2.50 (dotted line), 2.80 (solid line) and
3.00 (dashed line) as a function of -h.
\item{Fig. 3.} The order parameter $O$
as in Fig. 1 except on a $12^3$ lattice.
\item{Fig. 4.} The normalised susceptibility $\chi / V$
as a function of -h for lattices of the sizes $12^3$ (solid) and
$16^3$ (dashes).
\item{Fig. 5.} The location of the transition point $-h_{tr}$
as a function of the inverse volume for the lattice sizes
$24^3$, $16^3$ and $12^3$ corresponding to $\beta_4$ = 2.80.
The triangles are the respective points
and the line is the best fit.
\item{Fig. 6.} The locations of the physical points $-h_{ph}$ (circles)
are shown in relation to the transition points $-h_{tr}$ (triangles)
as a function of the three dimensional coupling $\beta$
for lattice size $L_s = 16$.
\vfill\eject
\end